\documentclass[
 reprint,
showpacs,preprintnumbers,
 amsmath,amssymb,
 aps,
prb,
]{revtex4-1}

\usepackage{graphicx}% Include figure files
\usepackage{dcolumn}% Align table columns on decimal point
\usepackage{bm}% bold math
\usepackage{color}
\usepackage{pstricks}
\usepackage[normalem]{ulem}
\usepackage{comment}

\def\be{\begin{equation}}
\def\ee{\end{equation}}
\def\q{k}

\def\connection{\mbox{\boldmath${\cal A}$}}
\def\curvature{\mbox{\boldmath${\Omega}$}}

\newcommand{\Ket}[1]{|#1\rangle}

\newcommand{\MatEl}[3]{\langle#1|#2|#3\rangle}
\newcommand{\Avg}[1]{\langle#1\rangle}
\newcommand{\bea}{\begin{eqnarray}}
\newcommand{\eea}{\end{eqnarray}}
\renewcommand{\vec}[1]{{\bf #1}}

\begin{document}

\title{Contrasting lattice geometry dependent versus independent quantities:\\  
  ramifications for Berry curvature, energy gaps, and dynamics}

\author{Steven H. Simon${}^1$}
\author{Mark Rudner${}^{2,3}$}
\affiliation{${}^1$ Rudolf Peierls Centre, Oxford University, OX1 3NP, United Kingdom}
\affiliation{${}^2$ Niels Bohr Institute, University of Copenhagen, 2100 Copenhagen, Denmark}
\affiliation{${}^3$ Center for Quantum Devices and Niels Bohr International Academy}
\date{\today}% It is always \today, today,

\begin{abstract}
  In the tight-binding description of electronic, photonic, or cold
  atomic dynamics in a periodic lattice potential, particle motion is
  described in terms of hopping amplitudes and potentials on an
  abstract network of discrete sites corresponding to physical
  orbitals in the lattice.  The physical attributes of the orbitals,
  including their locations in three-dimensional space, are
  independent pieces of information.  In this paper we identify a
  notion of geometry-independence: any physical quantity or observable
  that depends only on the tight-binding parameters (and not on the
  explicit information about the orbital geometry) is said to be
  ``geometry-independent.''  The band structure itself, and for
  example the Chern numbers of the bands in a two-dimensional system,
  are geometry-independent, while the Bloch-band Berry curvature is
  geometry-dependent.  Careful identification of geometry-dependent
  versus independent quantities can be used as a novel principle for
  constraining a variety of results.  By extending the notion of
  geometry-independence to certain classes of interacting systems,
  where the many-body energy gap is evidently geometry-independent, we
  shed new light on a hypothesized relation between many-body energy
  gaps of fractional Chern insulators and the uniformity of Bloch band
  Berry curvature in the Brillouin zone.  We furthermore explore the
  geometry-dependence of semiclassical wave packet dynamics, and use
  this principle to show how two different types of Hall response
  measurements may give markedly different results due to the fact
  that one is geometry-dependent, while the other is
  geometry-independent.  Similar considerations apply
  for %situation exists for
  anomalous thermal Hall response, in both electronic and spin
  systems.
\end{abstract}

\pacs{Valid PACS appear here}% PACS, the Physics and Astronomy
\maketitle

\section{Introduction}

There has been a recent explosion of interest in the topological properties of non-interacting electronic band structures~\cite{HasanKaneReview,QiZhangReview,FelserReview,BansilReview}. Of the quantities that have come to the fore during this time, the Berry connection and 
Berry curvature of the electron bands have played a particularly prominent role.  These quantities are not only of importance for understanding the topological nature of the band structure, but they also are crucial in determining the dynamics of particles within the band when electromagnetic perturbations are applied\cite{NiuReview}. 

Often when one studies band structure, one starts with a tight binding model which defines a set of orbitals 
$i,j,\ldots,$  along with potentials and hopping amplitudes $\{t_{ij}\}$ between them. 
These potentials and hopping amplitudes completely define a number of crucial properties of the system, such as its eigenenergies, 
as well as topological information such as the Chern numbers of its bands.  
A completely independent piece of information is the set of real-space positions $\{{\bf r}_i\}$ of the orbitals.    While the eigenenergies of the system 
are encoded entirely in the hoppings $\{t_{ij}\}$ and do not depend on the positions $\{{\bf r}_i\}$, quantities such as the Berry curvature do depend explicitly on these positions.
We say that the eigenenergies are {\it geometry independent} whereas the Berry curvature is {\it geometry dependent}.  
The purpose of this paper is to explore the ramifications of this notion of geometry dependence, in particular for
the Berry curvature.   

We will be able to apply some of these considerations not only to
noninteracting systems, but also to certain interacting models as
well.  In every case we will 
consider which quantities depend on geometry independent information,
such as hoppings between orbitals, and which quantities depend on
geometric information, such as the positions of orbitals.
Distinguishing the geometry independent from geometry dependent quantities turns out to give us some very useful and important 
insight into
physical systems and their responses to perturbations.

This paper is organized as follows.
In Sec.~\ref{sec:geomindep} we 
introduce the models that we consider and define the geometry dependent and geometry independent quantities.
In Sec.~\ref{sec:geom} we focus on the Berry curvature and we show that it is geometry dependent.  
In Sec.~\ref{sec:FCI} we provide our first example of  how focusing on geometry dependent versus geometry independent quantities can 
provide helpful new insight and results.   Specifically, we consider the widely discussed hypothesis that fractional Chern insulators \cite{Bergholtz,Para}, the lattice analogues of fractional quantum Hall systems, should have larger many-body gaps when the Berry curvature is as uniform (``flat'') as possible in the Brillouin zone.   
Since the Berry curvature depends on the detailed geometry of the lattice, and under certain conditions (such as pointlike Hubbard interactions) the many-body gaps do not, flat Berry curvature must not be required for obtaining large many-body gaps.   
Further discussion of this hypothesis and a potential refinement based on our considerations are given in Appendix~\ref{app:geometric}.

We then turn in Sec.~\ref{sec:semi} to study the dynamical equations
of motion of noninteracting particles in a partially filled band in an
applied electric field.  Since these equations involve the Berry
curvature, one might think that a change in geometry (change in the
positions of orbitals) may have an effect on the equations of motion.
On the other hand, since the tight-binding Hamiltonian defined by
$\{t_{ij}\}$ and the band structure are independent of the geometric
information in $\{\vec{r}_i\}$, one might expect the
dynamics 
to be completely unchanged by a deformation of $\{\vec{r}_i\}$ that
leaves $\{t_{ij}\}$ invariant.  In Sec.~\ref{sec:semi} we address this
issue; we show that the semiclassical wave packet dynamics are indeed
geometry dependent, and discuss the origins and implications of this
dependence.

Finally in Section~\ref{sec:hall} we turn our attention to 
electrical and thermal Hall transport for the case of a partially
filled band of noninteracting electrons.  For simplicity we focus on
the anomalous case, where no external magnetic field is applied but
the system internally breaks time reversal symmetry 
and thereby supports nonzero Hall transport
coefficients.  There are several different experimental configurations
which might be used to measure such transport coefficients, which at
first examination might be expected to give the same result.  On the
contrary we find some configurations give a geometry independent
result whereas others give a geometry dependent result.  For example,
if a physical system with noninteracting fermions has a chemical
potential difference applied between two contacts, we obtain a
geometry independent Hall response.  However, if we couple these
fermions to an external electric field and replace the chemical
potential difference with a uniform electric field, we instead find a
geometry dependent Hall conductivity.  We explain this phenomenon in
detail in Sec.~\ref{sec:hall}, and discuss how similar considerations can be applied to anomalous thermal Hall transport as well as thermal transport by magnons.
We give further
calculational details of the Hall response in the disorder-free limit in
Appendix \ref{app:current}.  Note that throughout this work we set
$\hbar = 1$.

\section{Geometry Independence}
\label{sec:geomindep}

In this section we define the notions of geometry dependence and independence that we explore in this paper.
Although it is not necessary to describe crystalline systems in terms of tight-binding models, %it is convenient to do so here.  
we will introduce these notions within the tight-binding framework where our results are simplest to discuss.
As we discuss in the conclusions, the ideas and results are more general. 

To describe electronic states in a crystalline system, we define an
orthonormal basis of orbitals
$\{\Ket{\alpha,\,\vec{R} + \vec{x}_\alpha}\}$, where the label
$\alpha$ runs over the different types of orbitals (including spin
indices) within the unit cell, $\vec{R}$ runs over all lattice
vectors, and $\vec{x}_\alpha$ is the position of orbital $\alpha$
within the unit cell (relative to the unit cell origin which we take to be at the lattice
point $\vec{R}$). %, and the absolute position of the orbital is given by ${\bf r}_i = {\bf R} + {\bf x}_\alpha$.
In the absence of
disorder and interactions, the single-particle band structure of the
system can be found from the translationally-invariant tight-binding
Hamiltonian
\begin{equation}
H_{0} = \sum_{ij} t_{ij} c^\dagger_i c_j   + \mbox{h.c.},
\label{eq:Hhopping}
\end{equation}
where $i$ and $j$ each run over all values of $\vec{R}$ and $\alpha$, and $c^\dagger_i$ and $c_j$ are creation and annihilation operators for electrons in the corresponding basis modes, respectively.
The absolute position of orbital $i$ is given by ${\bf r}_i = {\bf R} + {\bf x}_\alpha$, for the corresponding values of $\vec{R}$ and $\alpha$.  {These positions are sometimes known as the {\it orbital embeddings}\cite{HaldaneLecture,HaldaneUnpub,Stuckelberg}}.
The parameters $\{t_{ij}\}$ encode the information of the tight-binding Hamiltonian $H_0$ in a weighted graph (with ``on-site'' energies represented as the diagonal matrix elements $\{t_{ii}\}$).

Importantly, as we discuss and generalize below, once the
tight-binding Hamiltonian has been written in the form of
Eq.~(\ref{eq:Hhopping}), all of the physical information about the
basis orbitals $\{\Ket{\alpha,\,\vec{R} + \vec{x}_\alpha}\}$ has been
abstracted away.  In particular, the graph defined by the matrix
elements $\{t_{ij}\}$ does {\it not} explicitly encode the crystal
geometry defined by the positions $\{{\bf r}_i\}$  of the
orbitals.  Some physical properties are determined solely by the
parameters $\{t_{ij}\}$, and thus do not explicitly depend on the
crystal geometry.  We call such properties {\it geometry-independent}.
Other physical properties, as we will see, require knowledge of both
the tight-binding parameters and the physical information about the
nature of the basis orbitals; we refer to these properties as being
{\it geometry-dependent}.  {For simplicity, throughout this work we
  keep the {\it lattice geometry}, i.e., the unit cell size and shape
  (defined by the lattice vectors $\{\vec{R}\}$), fixed.}

A central example of a geometry-independent property is the single-particle band structure itself.
The electronic dispersion relation follows directly by diagonalizing $H_0$ in Eq.~(\ref{eq:Hhopping}), and thereby is determined solely by the values of the tight-binding parameters $\{t_{ij}\}$.
By diagonalizing $H_{0}$, we also obtain the coefficients of the corresponding eigenvectors $\{\Ket{\psi_n(\vec{k})}\}$ %(in the $\{\Ket{\alpha,\,\vec{R} + \vec{x}_\alpha}\}$ basis) 
directly from the parameters $\{t_{ij}\}$.
Here $n$ is a band index, and $\vec{k}$ is the wave vector (crystal momentum).
Expanding $\Ket{\psi_n(\vec{k})}$ in the orbital basis $\{\Ket{\alpha,\,\vec{R} + \vec{x}_\alpha}\}$ as 
\begin{equation}
\label{eq:Psi_decomp}\Ket{\psi_n(\vec{k})} = \sum_{\vec{R}, \alpha}  e^{i\vec{k}\cdot\vec{R}}\psi_{n\alpha}(\vec{k})\, \Ket{\alpha,\,\vec{R} + \vec{x}_\alpha},
\end{equation}
we see that the amplitudes $\psi_{n\alpha}(\vec{k})$ (or, equivalently, the eigenstate wave function amplitudes on all of the sites) are in fact geometry-independent, i.e., they are determined solely by $\{t_{ij}\}$ and do {\it not} depend explicitly on the orbital positions, $\{\vec{x}_\alpha\}$.

To see an example of a geometry-dependent quantity, we turn to Bloch's theorem, which is formulated in the coordinate space of the physical system (not simply on the graph $t_{ij}$).
According to Bloch's theorem, the single-particle eigenstates of an electron in the periodic potential of a crystal lattice 
can be decomposed in the form 
\begin{equation}
\label{eq:udef}
 |\psi_n({\bf \q}) \rangle = e^{i {\bf \q} \cdot \hat{\bf r}} |u_n({\bf \q}) \rangle,
\end{equation}
where $\hat{\vec{r}}$ is the position operator, $n$ is the band index, and $|u_n({\bf \q}) \rangle$ is the periodic Bloch function.
Importantly, in the position representation, $|u_n({\bf \q}) \rangle$ exhibits the same periodicity as the lattice.

Through the definition of the periodic Bloch function $\Ket{u_n(\vec{k})}$ in Eq.~(\ref{eq:udef}), the amplitudes $\{u_{n\alpha}(\vec{k})\}$ in 
\begin{equation}
\label{eq:Bloch_decomp} \Ket{u_n(\vec{k})} = \sum_{\vec{R}, \alpha} u_{n\alpha}(\vec{k}) \, \Ket{\alpha,\,\vec{R} + \vec{x}_\alpha}
\end{equation}
{\it must} depend explicitly on the positions $\{\vec{x}_\alpha\}$ of the orbitals; these amplitudes are therefore {\it geometry-dependent}.
Specifically, if we let $\hat{\vec{r}}\Ket{\alpha,\,\vec{R} + \vec{x}_\alpha} \approx (\vec{R} + \vec{x}_\alpha)\Ket{\alpha,\,\vec{R} + \vec{x}_\alpha}$, then Eqs.~(\ref{eq:Psi_decomp})-(\ref{eq:Bloch_decomp}) imply 
\begin{equation}
\label{eq:u_nk}u_{n\alpha}(\vec{k}) = e^{-i\vec{k}\cdot\vec{x}_\alpha}\psi_{n\alpha}(\vec{k}).
\end{equation}
Thus we see that, while $\psi_{n\alpha}(\vec{k})$ can be determined directly from $H_0$, the amplitudes $\{u_{n\alpha}(\vec{k})\}$ require knowledge both of the tight-binding matrix elements and the physical natures of the corresponding orbitals. %/basis states.

Note that in writing Eq.~(\ref{eq:u_nk}) via the approximation
$\hat{\vec{r}}\Ket{\alpha,\,\vec{R} + \vec{x}_\alpha} \approx (\vec{R}
+ \vec{x}_\alpha)\Ket{\alpha,\,\vec{R} + \vec{x}_\alpha}$, we consider
the orbitals to be pointlike in space; more generally, off diagonal
matrix elements
$\MatEl{\alpha',\,\vec{R}' +
  \vec{x}_{\alpha'}}{\hat{\vec{r}}}{\alpha,\,\vec{R} + \vec{x}_\alpha}
\neq 0$ may also appear.  For well-separated orbitals, these
off-diagonal matrix elements are expected to be small.  When multiple
orbitals on the same atom are involved, the off-diagonal contributions
may become more significant.  While the considerations below can be
extended to include these off-diagonal contributions in the Berry
connection, for simplicity we restrict our discussion to the limit of
pointlike orbitals where Eq.~(\ref{eq:u_nk}) holds.  These
contributions may change quantitative results, but do not change our
qualitative considerations and
conclusions.

Our goal in this work is to characterize the geometry dependence of various physical properties of crystalline systems, analogously to the discussion above. 
Of particular relevance, we will explore the geometry dependence of the Bloch band Berry curvature and corresponding consequences for transport.  For generality, we will extend the setting beyond the simple translation-invariant quadratic Hamiltonian $H_0$ in Eq.~(\ref{eq:Hhopping}).
In particular, we may consider the role of disorder\footnote{One may also consider disorder in the position of orbitals without changing any of the on-site potentials, the hoppings, or the interactions between orbitals.  This would be a purely geometric change which does not alter the Hamiltonian. However, as emphasized in section \ref{sec:semi} such a change in position would change the coupling to an externally applied electric field.} (both in hopping and potential), described in tight-binding form via
%Sometimes we may also add a disorder term 
\begin{equation}
 H_{\rm dis} = \sum_{ij} \delta t_{ij} c^\dagger_i c_j + h.c..
 \label{eq:Hdis}
\end{equation}
We may further consider a general extended Hubbard-like interaction term\footnote{More generally, the interaction may involve four orbitals, through terms of the form $\sum_{ijkl}V_{ijkl} c^\dagger_i c^\dagger_j c_k c_l$. However, for our discussion of geometry-independent properties, the Hubbard-like interaction that we consider in Eq.~(\ref{eq:Hint}) is most relevant.}
\begin{equation}
H_{\rm int} =  \sum_{i,j} U_{ij} n_i n_j
\label{eq:Hint}
\end{equation}
between electrons in orbitals $i$ and $j$, where $n_i = c^\dagger_i c_i$ is the number of particles in orbital $i$.
As discussed above for $H_0$, the geometry of the crystal is abstracted away from the parameters $\{\delta t_{ij}\}$ and $\{U_{ij}\}$; we  therefore extend the definition of geometry independence to characterize any physical quantities that are determined solely by the values of $\{t_{ij}, \delta t_{ij}, U_{ij}\}$ with no explicit input about the geometry or physical nature of the corresponding orbitals/basis states.

\section{Geometric Transformation of Berry Curvature}
\label{sec:geom}

In this section we discuss the geometry dependence of the Bloch band Berry curvature by examining its explicit dependence on the positions $\{\vec{x}_\alpha\}$ of the orbitals as introduced in Sec.~\ref{sec:geomindep}.  For now we consider the case of translation-invariant, non-interacting systems, with $H_{\rm dis} = 0$ and $H_{\rm int} = 0$.
Using the periodic Bloch functions $\{\Ket{u_n(\vec{k})}\}$ defined via Eqs.~(\ref{eq:udef}) and (\ref{eq:Bloch_decomp}), we define the Berry connection of the $n^{\rm th}$ band as
\begin{equation}
\label{eq:connection}
 \connection_n({\bf \q}) = i \langle u_n({\bf \q}) | \nabla_{\bf \q} | u_n({\bf \q}) \rangle.
\end{equation}
The corresponding Berry curvature is given by\cite{NiuReview}:
\begin{equation}
\label{eq:curvature}
\curvature_{n}({\bf \q}) = \nabla_{\bf \q} \times \connection_n({\bf \q}).
\end{equation}

We now investigate the geometry dependence of $\connection_n({\bf \q})$ and $\curvature_n(\vec{k})$.
Consider two systems, (1) and (2), described by the same tight-binding parameters $\{t_{ij}\}$ in Hamiltonian $H_0$, in Eq.~(\ref{eq:Hhopping}).
The physical details of the systems may be different, however, with orbitals $\{\alpha\}$ being of possibly different types, at positions $\{\vec{x}^{(1)}_\alpha\}$ and $\{\vec{x}^{(2)}_\alpha\}$.
As discussed in the previous section, systems (1) and (2) will exhibit identical band structures.
As we now show, the Berry connections and Berry curvatures in general depend on the details of $\{\vec{x}^{(1)}_\alpha\}$ and $\{\vec{x}^{(2)}_\alpha\}$.

To characterize the relationship between the Berry connections and curvatures of systems (1) and (2), for each orbital $\alpha$ we define the relative displacement %$\delta\vec{x}_\alpha = \vec{x}^{(2)}_\alpha - \vec{x}^{(1)}_\alpha$, 
\begin{equation} \label{eq:move1}
  \delta\vec{x}_\alpha \equiv \vec{x}^{(2)}_\alpha - \vec{x}^{(1)}_\alpha
\end{equation}
to characterize the relative shift of orbital $\alpha$ between the two structures. 
As argued above, the wave function amplitudes $\{\psi_{n\alpha}(\vec{k})\}$ in Eq.~(\ref{eq:Psi_decomp}) must be the same for systems (1) and (2).
However, according to Eq.~(\ref{eq:u_nk}), within the approximation of pointlike orbitals, the amplitudes $u^{(1)}_{n\alpha}({\bf \q})$ and $u^{(2)}_{n\alpha}({\bf \q})$ of the corresponding periodic Bloch functions must be related by 
\begin{equation}
\label{eq:move2}
u^{(2)}_{n\alpha}({\bf \q})  = e^{-i {\bf \q} \cdot {\bf \delta x_{\alpha}}} u^{(1)}_{n\alpha}({\bf \q}).  
\end{equation}
In principle there could also be a relative gauge transformation $e^{i \chi({\bf k})}$ on the right hand side of Eq.~(\ref{eq:move2}), where $\chi(\vec{k})$ is an ($\alpha$-independent) arbitrary single-valued function over the Brillouin zone.  
Here we ignore this possibility; we will further discuss this particular gauge choice near Eq.~(\ref{eq:rshift}).
The corresponding Berry connections are related via %
\begin{equation} 
  \label{eq:A21}\connection^{(2)}_n({\bf \q}) = \connection^{(1)}_n({\bf \q}) +  \overline{\delta\vec{x}_n}({\bf k}),
\end{equation} 
where 
\begin{equation}  
  \overline{\delta\vec{x}_n}({\bf k}) \equiv \sum_\alpha {\bf  \delta x}_{\alpha} |u_{n\alpha}({\bf \q})|^2. \label{eq:dxdef}
\end{equation}
Here $\overline{\delta\vec{x}_n}(\vec{k})$ may be interpreted as a $\vec{k}$-dependent shift of the electron position
within the unit cell.
The corresponding Berry curvature then becomes
\begin{equation}
\label{eq:omshift}
\curvature^{(2)}_n({\bf \q})  = \curvature_n^{(1)}({\bf \q})  +  \nabla_{\bf \q} \times \overline{\delta\vec{x}_n}({\bf k}).
\end{equation}
This relation has been long known to experts (see, for example, Ref.~\onlinecite{Milo,Stuckelberg} and the supplementary material of Ref.~\onlinecite{Jackson2015}).   {As discussed in Refs.~\onlinecite{CooperRMP,Stuckelberg, Fruchart}, this type of transformation on the orbital positions corresponds to a unitary transformation on the Bloch Hamiltonian $H({\bf k})$.} 

We note that topological properties, such as the Chern number $C_n$ of band $n$ in a two-dimensional system, are geometry {\it independent}: through its definition as an integral over the entire (periodic) two-dimensional Brillouin zone (BZ), 
$$
C_n = \frac{1}{2 \pi} \int_{\rm BZ} d^2 \q \,\, \Omega_n({\bf \q}),
$$
we see that the contribution of the second term in Eq.~(\ref{eq:omshift}) vanishes by Stokes' theorem\footnote{Stokes' theorem applies here since $\overline{\delta\vec{x}_n}({\bf k})$ is continuous, single valued, and nonsingular over the Brillouin zone.}.
Similarly, for three dimensional systems the Berry flux through a fixed surface in $k$-space,
$$
 C = \frac{1}{2 \pi} \int {\bf dS} \cdot {\curvature}_n({\bf k}),
$$
is geometry independent.  
Thus, for example, if there is a Weyl node carrying a monopole of Berry flux, this flux must be the same for any systems with identical tight-binding parameters, independent of the geometric information about atomic coordinates.  Note, however, that an integral of the Berry curvature over a domain with boundary is not generally geometry independent. 

\section{Geometric Stability Hypothesis}
\label{sec:FCI}

In the many-body context, bands with nontrivial Berry curvature play a
central role for example in the fractional quantum Hall effect, which
occurs when Landau levels are partially filled with interacting
particles.  Landau levels, which are dispersionless and feature a
uniform Berry curvature throughout the Brillouin zone, are obtained
when a free electron is subjected to a uniform magnetic field.  In
recent years, considerable interest has grown in the study of
analogous phases known as fractional Chern insulators\cite{KolRead,
  Demler, Hafezi, MollerCooper2009,  Neupert2011,
  Tang2011, Sun, Regnault, Bernevig, Bergholtz, Para}, which may arise in partially filled,
topologically-nontrivial bands of interacting particles on a two
dimensional lattice.  By analogy to the Landau levels that underlie
the fractional quantum Hall effect, it has long been hypothesized that
stable (i.e., large gap) fractionally filled states on the lattice
most favorably arise when the Berry curvature is uniform in the
Brillouin zone\cite{Tang2011,
  Sun,Sheng,Bernevig,Roy,Para,Bergholtz,Udagawa,Bauer,Jackson2015}.  In a
slightly more general form (see Ref.~\onlinecite{Jackson2015} and
Appendix \ref{app:geometric}), this hypothesis has been called a
``geometric stability hypothesis.''  As we now discuss (see also
Ref.~\onlinecite{Milo}), based on the notions of geometry dependence
and independence formulated above, this hypothesis is not well-defined
as stated.

Note that, as our aim in this paper is to highlight the notions of geometry dependence and independence defined in Sec.~\ref{sec:geomindep} and to demonstrate how they may be used obtain new perspectives on various physical phenomena, an extended discussion of the geometric stability hypothesis is beyond our scope. 
However, for context, in Appendix \ref{app:geometric} we provide a more detailed summary of the proposed conjectured relations between many-body gaps and quantities that characterize the geometry of a system's Bloch bands.

Consider a two-dimensional system with a Hamiltonian as defined in Eqs.~(\ref{eq:Hhopping}) and (\ref{eq:Hint}), and partial filling of the highest occupied band. 
For example, we may consider a nearest neighbor hopping Hamiltonian on some lattice, with an on-site Hubbard interaction. 
With these definitions, the Hamiltonian is 
geometry independent in the sense defined in Sec.~\ref{sec:geomindep} (making no reference to the positions of orbitals).
Hence, the many-body gap is entirely independent of changes in geometry %as given by Eq.~(\ref{eq:move1}) 
so long as the potentials, hopping matrix elements, and interactions are kept fixed.
However, under this change in geometry the Berry curvature (and hence its flatness over the Brillouin zone) changes according to Eqs.~(\ref{eq:move1}) and (\ref{eq:omshift}).  

As the arguments above show, the flatness (or uniformity) of the Berry curvature can be tuned independently of the many-body gap; this seemingly 
contradicts the idea of a simple geometric stability hypothesis\cite{Milo, note:Milo}. 
Indeed this conundrum was also noted in the supplementary material of Ref.~\onlinecite{Jackson2015}, where it was suggested that certain positions in the unit cell are more physical than others.  
On the other hand, one may justifiably study many different physical systems with different geometries (yet equal many-body gaps), with none being particularly more valid than any other. 
In Appendix \ref{app:geometric} we offer an alternative formulation of the conjecture on the stability of fractional Chern insulator phases that takes into account the special role of the geometries identified in Ref.~\onlinecite{Jackson2015} and in numerical studies~\cite{Bernevig, Jackson2015, Bauer}, as well as the notions of geometry-dependence formulated in the present work.
We emphasize that a further investigation of the stability of fractional Chern insulators is beyond the scope of this work; our intention here is to highlight the utility of the notion of geometry-independence and to use it to suggest a direction for future work.

\section{Semiclassical Dynamics}
\label{sec:semi}

In this section we consider the semiclassical dynamics of wave packets for noninteracting particles.   It is well known that Berry curvature, which is geometry dependent as shown in Eq.~(\ref{eq:omshift}), is relevant to semiclassical dynamics.  Here we seek to elucidate the origin of this geometry dependence.   In this section we will consider a geometry independent hopping Hamiltonian as in Eq.~(\ref{eq:Hhopping}), and neglect interparticle interactions. %we set the interaction to zero.   
Crucially, to determine the equations of motion, we must allow the particles in the model to also be coupled to a uniform externally applied electric field ${\bf E}({\bf r}) = -\nabla \phi({\bf r})$.  The coupling to the externally applied potential, given the assumption of pointlike orbitals (see Sec.~\ref{sec:geomindep}), can be written as 
\begin{equation}
 H_{\rm ext} = \sum_{i} \phi({\bf r}_i) n_i,
\label{eq:Hext}
\end{equation}
where $n_i = c^\dagger_i c_i$ is the occupation number operator for orbital $i$ and ${\bf r}_i$ is the position of orbital $i$.    Note that this coupling  depends explicitly on the geometry --- in particular it depends on the orbitals' positions.

\subsection{Berry Curvature and Semiclassical Dynamics}
\label{sub:berrysemi1}

Considering wave packets in a Bloch band structure, one can determine a set of semiclassical equations of motion.  The full derivation of such equations (and extensive discussion) is given in Refs.~\onlinecite{NiuReview,ChangNiu96} for example.   
Here we will repeat just a few key pieces of the discussion. 

Using a general 
  construction (see, e.g., Sec.~IV.A of Ref.~\onlinecite{NiuReview}), 
we form a wave packet 
$\Ket{W(t)}$ in band $n$ as
\be
\label{eq:W} \Ket{W(t)} = \int d{\bf \q} \,\, w({\bf \q}, t) |\psi_n({\bf \q}) \rangle,
%\Ket{W(\vec{r}_c, \vec{\q}_c)} = \int d{\bf \q} \,\, w({\bf \q}; \vec{r}_c,\vec{\q}_c) |\psi_n({\bf \q}) \rangle,
\ee
where $|\psi_n({\bf \q})\rangle$ is the full Bloch wave function for the $n^{th}$ band (including the plane wave part), see Eq.~(\ref{eq:Psi_decomp}).    %This wave packet is constructed to be %around position ${\bf r}_c$ in real space and around wave vector ${\bf \q}_c$ in reciprocal space.    Here  ${\bf r}_c$ and  ${\bf \q}_c$ are parameters, which can be a functions of time, and the amplitudes $w({\bf \q}; \vec{r}_c,\vec{\q}_c)$ %is a distribution which 
The amplitudes $w(\vec{k}, t)$ describe the (time-dependent) structure of the wave packet.   
We consider wave packets that are tightly localized in reciprocal space, such that %the probabilities
  $|w(\vec{k},t)|^2$ %$|w({\bf \q}; \vec{r}_c,\vec{\q}_c)|^2$  %are
  is approximately distributed like a delta-function centered at  wave vector ${\bf \q}_c$ (which may be a function of time),  
\begin{equation}
  \label{eq:kc} {\bf \q}_c(t) = \int d{\bf \q} \,\, {\bf \q} \,\,|w({\bf \q}, t)|^2.  %|w({\bf \q}; \vec{r}_c,\vec{\q}_c)|^2. 
\end{equation}
The real space position of the wave packet is determined by the phases of the complex amplitudes $w(\vec{k},t)$, %$w({\bf \q}; \vec{r}_c,\vec{\q}_c)$,
as well as the structure of the wave function components $|\psi_n({\bf \q}) \rangle$ and the orbital positions $\{\vec{x}_\alpha\}$.
As discussed in Refs.~\onlinecite{NiuReview,ChangNiu96}, the average position of the wave packet can be written as 
\begin{equation}
 {\bf r}_c(t) =  {\bf F}[w({\bf \q}, t)]+ \connection_n({\bf \q}_c), %{\bf F}[w({\bf \q}; \vec{r}_c,\vec{\q}_c)]+ \connection_n({\bf \q}_c),
 \label{eq:rc}
\end{equation}
where $\connection_n({\bf \q}_c)$ is the Berry connection for the band [Eq.~(\ref{eq:connection})] and (suppressing time arguments)
\be \label{eq:F}
 {\bf F}[w({\bf \q})] =  \int d{\bf \q}  \,\, w^*({\bf \q}) \, (i \nabla_{\bf \q}) \,w({\bf \q}) = -\nabla_{\bf \q} \arg w({\bf \q}) \Big\vert_{{\bf \q} = {\bf \q}_c}.
\ee
From here on we will drop the subscript ${c}$, such that ${\bf r}$ and ${\bf k}$ will describe the center of mass position and wave vector of the wave packet.

As discussed in Ref.~\onlinecite{NiuReview}, in the presence of a weak electric field, the center of mass position and momentum of the wave packet in band $n$ (assuming non-degenerate bands\footnote{In the case of degenerate bands, the equations of motion allow transitions between the degenerate states\cite{NiuReview}.  If bands are near degenerate then the non-degenerate equations of motion are applicable only for perturbations weak enough not to mix the states}) evolve according to the semiclassical equations of motion 
\begin{eqnarray} \label{eq:motion1a}
\dot {\bf r} &=& \nabla_{\bf k} \varepsilon_n({\bf k}) - \dot {\bf k} \times {\bf \Omega}_n({\bf k}) \\ \label{eq:motion2a}
\dot {\bf k} &=& -e {\bf E},
\end{eqnarray}
where $\varepsilon_n({\bf k})$ and ${\bf \Omega}_n(\vec{k})$ are the band structure dispersion and Berry curvature [see Eq.~(\ref{eq:curvature})] for band $n$, and ${\bf E}$ is the externally applied electric field.  Wave packet dynamics governed by these equations have been investigated recently in cold atoms\cite{Price2012, Jotzu2014, Aidelsburger, Flaschner2016, Wimmer2017, Wintersperger2020}.
Note that we have 
assumed zero external magnetic field, 
although we will generally assume that time reversal symmetry is broken so that one can have a nonzero (anomalous) Hall response. 
Below we will drop the band-index subscript $n$ on $\varepsilon_n(\vec{k})$ and ${\bf \Omega}_n(\vec{k})$.

\subsection{Role of orbital positions in the unit cell}
\label{sub:changing}

We are now ready to compare the equations of motion for wave packets in two systems (1) and (2), described by the same tight-binding parameters $\{t_{ij}\}$ but different physical geometries (i.e., atomic positions $\{\vec{r}_i\}$), as discussed in Sec.~\ref{sec:geom}.
Given that the tight-binding Hamiltonians for systems (1) and (2) are identical, we may pick gauges for each such that the wave function amplitudes of the eigenstates $\psi_{n\alpha}^{(1)}(\vec{k})$ and $\psi_{n\alpha}^{(2)}(\vec{k})$ are identical in the site basis.
With this choice of gauge, the wave packets formed according to Eq.~(\ref{eq:W}) with identical choices of $w(\vec{k})$ will be spatially shifted with respect to one another.
{Note that specifying a relative gauge for the two systems is necessary in order to make physically-meaningful comparisons between wave packets in the two systems. The conclusions below about the {\it equations of motion} for wave packets in the two systems are gauge-independent.}

Using Eq.~(\ref{eq:A21}) in Eq.~(\ref{eq:rc}), we find the relative shift of center of mass position of the wave packet %in the $n^{th}$ band 
between systems (1) and (2): 
\begin{equation}
     {\bf r}^{(2)} ={\bf r}^{(1)}  +   \overline{ \delta {\bf x}}({\bf k}),
 \label{eq:rshift}
\end{equation}
where we have dropped the subscript $n$ on $\overline{ \delta {\bf x}}({\bf k})$ from Eq.~(\ref{eq:A21}) since we have focused on a particular band.
This expression has a natural physical interpretation in terms of the shift of the electronic wave function in the unit cell, see text below Eq.~(\ref{eq:dxdef}).

The equations of motion for wave packets in systems (1) and (2) are given by Eqs.~(\ref{eq:motion1a}) and (\ref{eq:motion2a}).
Note that Eq.~(\ref{eq:motion2a}) is identical for the two cases, as it makes no reference to ${\bf r}$ or to the Berry curvature.
On the other hand we obtain two different equations of motion from Eq.~(\ref{eq:motion1a}):
\begin{equation} \label{eq:motion1aold}
\dot {\bf r}^{(1)} = \nabla_{\bf k} \varepsilon({\bf k}) - \dot {\bf k} \times {\bf \Omega}^{(1)}({\bf k}) 
\end{equation}
and 
\begin{equation} \label{eq:motion1anew}
\dot {\bf r}^{(2)} = \nabla_{\bf k} \varepsilon({\bf k}) - \dot {\bf k} \times {\bf \Omega}^{(2)}({\bf k}),
\end{equation}
where the Berry curvatures ${\bf \Omega}^{(1)}(\vec{k})$ and ${\bf \Omega}^{(2)}(\vec{k})$ are related by Eq.~(\ref{eq:omshift}).

It is natural to ask whether these two equations describe the same
dynamics or not.  Since the two systems (in the absence of the applied
field) are described by identical tight-binding Hamiltonians, one
might expect the same dynamical response. Of course the
  shift in position of the orbitals should result in some ``trivial''
  changes in observables due to the change in location where the
  electrons reside.
  Beyond this trivial effect, one may ask whether the populations on the {\it sites}, $\Avg{n_i}$, 
  evolve identically in the two cases.
If the two sets of equations of motion were to describe the same dynamics in this latter sense (i.e., with a simple change of variables to account for the orbital shifts), one would be able to recover Eq.~(\ref{eq:motion1aold}) from Eq.~(\ref{eq:motion1anew}) by directly substituting the expressions
for ${\bf r}^{(2)}$ in terms of ${\bf r}^{(1)}$
[Eq.~(\ref{eq:rshift})] and ${\bf \Omega}^{(2)}$ in terms of
${\bf \Omega}^{(1)}$ [Eq.~(\ref{eq:omshift})] into
Eq.~(\ref{eq:motion1anew}).  As we now show, performing this
substitution does not return Eq.~(\ref{eq:motion1aold}); hence
Eqs.~(\ref{eq:motion1aold}) and (\ref{eq:motion1anew}) describe
different physical dynamics.

To track down the source of this difference, we perform the substitution described above and obtain: 
$$
 \frac{d}{d t} \! \left[{\bf r}^{(1)} + %\rule[0pt]{0pt}{20pt}
\overline {\bf 
\delta x} ({\bf k}) \right]  = \nabla_{
\bf k} \varepsilon({\bf k}) - \dot {\bf k} \times \!\! \left[ {\bf \Omega}^{(1)}({\bf k})\! + \! \nabla_{\bf k} \! \times  \! \overline {\bf 
\delta x} ({\bf k}) %\rule[0pt]{0pt}{20pt} 
\right].
$$
Evaluating  the left hand side of this expression gives $\frac{d}{d t} \! \left[{\bf r}^{(1)} + \overline {\bf \delta x} ({\bf k}) \right]  = \dot {\bf r}^{(1)} + (\dot{\vec{k}}\! \cdot\! \nabla_{\vec{k}})\,  \overline {\bf \delta x} ({\bf k})$. 
Regrouping terms, we obtain
\begin{equation}
\dot {\bf r}^{(1)} =  \nabla_{
\bf k} \varepsilon({\bf k}) - \dot {\bf k} \times {\bf \Omega}^{(1)}({\bf k}) + \dot {\bf r}_{\rm extra},
\label{eq:final1}
\end{equation}
where 
\begin{equation}
\dot{\vec{r}}_{\rm extra} = - \dot{\vec{k}} \times \left[ \nabla_{\bf k} \! \times  \! \overline {\bf 
\delta x} ({\bf k})\right] - (\dot{\vec{k}} \cdot \nabla_{\vec{k}})\,  \overline {\bf \delta x} ({\bf k}).
%\dot r_{Extra}^a = -\epsilon^{abc} \dot k^b \epsilon^{cde} \partial_{k^d} \overline { 
%\delta x^e} ({\bf k}) - \dot k^b \partial_{k^b} \overline { 
%\delta x^a} ({\bf k})
\label{eq:Extra}
\end{equation}
Thus we see that the 
motion of a wave packet in system (2) 
differs from that of a wave packet in system (1) 
by the additional term $\dot {\bf r}_{\rm extra}$ appearing in Eqs.~(\ref{eq:final1}) and (\ref{eq:Extra}).

To determine the meaning of this additional term, it is worth simplifying the expression for $\dot {\bf r}_{\rm extra}$.  
Using Eq.~(\ref{eq:motion2a}) to make the replacement $\dot{\vec{k}} = - e\vec{E}$, along with the vector triple product identity $\vec{A} \times [\vec{B} \times \vec{C}] = \vec{B} (\vec{A}\cdot\vec{C}) - (\vec{A}\cdot\vec{B}) \vec{C}$, % for vectors $\vec{A} = \dot{\vec{k}}$, $\vec{B} = \nabla_{\vec{k}}$, and $\vec{C} = \overline {\bf \delta x} ({\bf k})$, 
we obtain
\begin{equation}
 \dot {\bf r}_{\rm extra} = \nabla_{\bf k} \left[ e {\bf E} \cdot \overline {\bf \delta x} ({\bf k})\right].  \label{eq:rdotres}
  \end{equation}

What is the origin of the extra term, $\dot{\vec{r}}_{\rm extra}$?  
The key is to note that the coupling of the electron to the uniform external electric field is sensitive to the positions of the orbitals.
In terms of the corresponding electric potential, we have ${\bf E} = -\nabla_{\bf r} \phi({\bf r}) $. 
The electrostatic potential for the wave packet in system (2) is given by  
\begin{eqnarray*}
 \phi\big({\bf r}^{(2)}\big)  &=& \phi\big({\bf r}^{(1)} + \overline { \bf \delta x} ({\bf k}) \big) \\ 
&\approx&  \phi\big({\bf r}^{(1)}\big)  - {\bf E} \cdot\overline {\bf \delta x} ({\bf k}).
\end{eqnarray*}
The energy of the wave packet in system (2) is therefore shifted by a $\vec{k}$-dependent energy $\delta \varepsilon({\bf k})$ relative to that of the corresponding wave packet in system (1):
\begin{equation}
  \delta \varepsilon({\bf k})  = -e \left[\phi\big({\bf r}^{(2)}\big) - \phi\big({\bf r}^{(1)}\big)\right] =   e {\bf E} \cdot \overline {\bf \delta x} ({\bf k}) \label{eq:deps}.
\end{equation}

In terms of wave packet propagation, a $\vec{k}$-dependent energy
shift $\delta \varepsilon({\bf k})$ essentially modifies the
dispersion; the corresponding change of group velocity is reflected in
the additional contribution to the velocity,
$\dot {\bf r}_{\rm extra}$ in Eq.~(\ref{eq:rdotres}).  The difference
in the equations of motion for wave packets in the two systems arise
from the fact that the electrons in the two systems couple differently
to the external field.   At
the level of the tight-binding Hamiltonian, we note that this
difference arises from the fact that the electric potential
$\phi(\vec{r})$ corresponding to the uniform electric field introduces
on-site energies that explicitly depend on the atomic coordinates
$\{\vec{r}_i\}$, thus breaking the geometry-independence of the
tight-binding Hamiltonian. {Nonetheless, as also pointed out in Ref.~\onlinecite{CooperRMP},
  Appendix D (and shown explicitly there for a two band model), both
  before and after the transformation, the system properly satisfies
  the semiclassical equations of motion, Eqs.~(\ref{eq:motion1a}) and (\ref{eq:motion2a}).}

\subsection{Lagrangian formulation}

For further insight into the geometry dependence of the semiclassical wave packet dynamics, it is instructive to investigate the 
role of the orbital positions $\{\vec{r}_i\}$ 
  in the semiclassical phase space Lagrangian.  
We begin with the Lagrangian describing the dynamics of a wave packet in band $n$ (in the absence of an applied magnetic field), as given in Eq.~(5.7) in Ref.~\onlinecite{NiuReview}:
\be\label{eq:Lagrangian}
L(\vec{r}, \vec{k}, \dot{\vec{r}}, \dot{\vec{k}}) =  \vec{k} \cdot \dot{\vec{r}} - \varepsilon_n(\vec{k}) + e\phi(\vec{r}) +\dot{\vec{k}} \cdot \boldsymbol{\mathcal{A}}_n(\vec{k}).
\ee
Note that here $\vec{r}$ and $\vec{k}$ correspond to $\vec{r}_c$ and $\vec{k}_c$, the average position and  momentum of the wave packet, see Eqs.~(\ref{eq:kc}) and (\ref{eq:rc}).  
Although $\vec{r}_c$ implicitly depends on $\vec{k}_c$ via Eq.~(\ref{eq:rc}), $\vec{r}$ and $\vec{k}$ should be treated as independent variables in the Lagrangian [they are in fact independent implicit parameters in the wave function $\Ket{W}$ in Eq.~(\ref{eq:W})]. 
In particular, $\partial_{\vec{k}} \phi(\vec{r}) = 0$.

Again comparing the behavior for two systems with identical tight-binding parameters but different atomic positions,  the Lagrangian in Eq.~(\ref{eq:Lagrangian}) can be written  as $L^{(s)} =  \vec{k}^{(s)} \cdot \dot{\vec{r}}^{(s)} - \varepsilon_n\big(\vec{k}^{(s)}\big) + e\phi(\vec{r}^{(s)}) +\dot{\vec{k}}^{(s)} \cdot \boldsymbol{\mathcal{A}}^{(s)}_n\big(\vec{k}^{(s)}\big)$ for system $s = 1,2$.
Using the coordinate transformation for $\vec{r}^{(2)}$ in terms of $\vec{r}^{(1)}$ given in Eq.~(\ref{eq:rshift}), along with its time-derivative $\dot{\vec{r}}^{(2)} = \dot{\vec{r}}^{(1)} + \frac{d}{dt}\overline{\delta  {\bf x}} ({\vec{k}})$ and the relation between Berry connections in Eq.~(\ref{eq:A21}), we express the Lagrangian for system (2) in terms of the coordinates for system (1):
\be
\label{eq:LagrangianOld}
L^{(2)} =  \vec{k}\cdot\dot{\vec{r}}^{(1)} - \varepsilon(\vec{k}) + e\phi\big(\vec{r}^{(1)} + \overline{\delta  {\bf x}} ({\vec{k}})\big) +  \dot{\vec{k}} \cdot \boldsymbol{\mathcal{A}}^{(1)}(\vec{k}) + {\rm T.D.}, %\frac{d}{dt}(\vec{k}\cdot\Avg{\delta \vec{x}}_{\vec{k}}).
\ee
where ``T.D.'' stands for ``total derivative.''
In writing Eq.~(\ref{eq:LagrangianOld}) we have used $\vec{k}^{(2)} = \vec{k}^{(1)} \equiv \vec{k}$ and $\dot{\vec{k}}^{(2)} = \dot{\vec{k}}^{(1)} \equiv \dot{\vec{k}}$, along with
$$\frac{d}{dt}\left[\vec{k}\cdot\overline{\delta  {\bf x}} ({\vec{k}})\right] = \vec{k}\cdot\frac{d}{dt}\overline{\delta  {\bf x}} ({\vec{k}}) + \dot{\vec{k}}\cdot\overline{\delta  {\bf x}} ({\vec{k}}).$$

Dropping the total derivative term, which does not affect the equations of motion, we see that the Lagrangian in Eq.~(\ref{eq:LagrangianOld}) takes the usual form as that in Eq.~(\ref{eq:Lagrangian}), with one twist: the potential is evaluated at a new $\vec{k}$-dependent position.
Although ordinarily we do not take $\vec{k}$-derivatives of the potential (see discussion above), due to the coordinate transformation that we have taken (where $\vec{r}^{(1)}$ does not correspond to the center of the wave packet, and therefore $\dot{\vec{r}}^{(1)}$ does not correspond to the wave packet's velocity), we will obtain $-\nabla\phi\cdot\overline{\delta  {\bf x}} ({\vec{k}}) = \vec{E}\cdot \overline{\delta  {\bf x}} ({\vec{k}})$ as a correction to the dispersion.
This term simply makes up for the bad parametrization of the wave function that we took by using a ``coordinate'' appropriate for a system with a different geometry, that doesn't actually correspond to the center of the wave packet in system (2).

\section{Anomalous  Hall Transport} 
\label{sec:hall}

Having highlighted 
the importance of the geometry-independent versus geometry-dependent
contributions to semiclassical dynamics, we now turn to examine the
Hall effect as an example.  We will consider both the electric current
response (the usual Hall effect) and the thermal response (the thermal
Hall effect, or Righi-Leduc effect). 
We will assume that the hopping between orbitals generically breaks
time reversal.  The appearances of nonzero Hall transport coefficients
in the absence of an external magnetic field are known as anomalous
Hall effects\cite{AHERMP}. Our considerations also directly apply
  to systems with a commensurate externally applied magnetic field,
  where we absorb the effects of the magnetic field into hopping
  phases and work with the magnetic unit cell as the elementary
  unit cell of the system.

{We emphasize that in the case of a filled band, for
  noninteracting (or weakly interacting) fermions, the electrical Hall
  conductance (and conductivity) will be quantized and is given by
  $ C e^2/h$ with $C$ the Chern number.  As mentioned in Sec.~\ref{sec:geom}, the Chern number is geometry independent. Similarly
  the Righi-Leduc effect will be quantized in this case.  The more
  interesting case to study, which we will focus on here, is the case
  of a partially filled band.}

The study of anomalous Hall responses is complicated, with several
distinct physical processes contributing, and a fair amount
of
competing claims persisting in the literature (see
Refs.~\onlinecite{AHERMP,Sinitsyn2007} for a detailed discussion of
the competing factors and the subtleties that have
  historically been the source of confusion). 
Because of these complexities it is useful to focus on the simple case of a disorder-free system, which clearly demonstrates the distinction between geometry dependent versus geometry independent responses.   Although this example is particularly simple to discuss, our main conclusions are not restricted to the disorder-free case.

\subsection{Electric Hall Current Response}

Consider a noninteracting system described by a tight-binding model as
defined in Eq.~(\ref{eq:Hhopping}).  Below we describe two scenarios
for measuring the electric Hall current response (current transverse
to the applied bias): ($i$) %in two different ways --- either
by applying a chemical potential difference between two contacts or
($ii$) by applying a uniform electric field.  We assume no coupling to
phonons.

% {Note that the simple models we consider here
 % do not provide any mechanism for the {electronic distribution to re-equilibrate.}} {\bf [MR: In the geometry independent case we do allow Hubbard interactions plus disorder. Does this count for equilibration? Are we selling ourselves short?]}

\subsubsection{Geometry Independent Response}
\label{subsub:electricindep}

Here we consider neutral particles, and imagine attaching the
system 
to two reservoirs which have a chemical potential difference between
them.  Taking a cylindrical geometry (periodic boundary conditions
in the direction perpendicular to the axis connecting the two
reservoirs),  we consider the total current that flows around the
circumference of the cylinder (i.e., perpendicular to the direction
between these reservoirs).

This model is completely geometry independent in the sense defined in
Sec.~\ref{sec:geom}. Nowhere in the Hamiltonian or the driving
perturbation do the positions of orbitals in the unit cell explicitly
enter.  Further, we may add geometry-independent disorder (i.e., a
random potential in the site basis) and interactions as in
Eqs.~(\ref{eq:Hdis}) and (\ref{eq:Hint}) and the Hamiltonian still
makes no reference to the positions of orbitals.  Finally, the
observable of interest, i.e., the current that flows around the
cylinder, can also be formulated as a geometry independent
quantity\footnote{One can imagine making a cut down the long axis of
  the cylinder and keeping track of the number of electrons per second
  that jump across this cut.  If we change the geometry of the system,
  the number of electrons per second jumping across the cut is
  completely unchanged so long as we do not move any orbitals across
  the cut.  Further, in the DC limit, due to current conservation, the current must be fully independent of where
  we make the cut.}.  Thus the response of this
system is
geometry-independent.   For reference, in Appendix
\ref{app:chemclean} we provide an explicit calculation of the current density
in the clean limit, and demonstrate that it is indeed
geometry-independent.

\subsubsection{Geometry Dependent Response}
\label{subsub:electricdep}

Instead of applying a chemical potential difference between
contacts, here we consider coupling the system to a uniform external electric field as
in Eq.~(\ref{eq:Hext}).  This perturbation is explicitly geometry-dependent; as shown in Sec.~\ref{sub:changing}, the geometry
dependence is crucial for obtaining the correct semiclassical
equations of motion.

To be concrete, and for maximal simplicity, we imagine a torus
geometry with an electric field directed around one of the
handles. This {situation} can be made more realistic by considering an
annulus {(Corbino)} geometry with electric field directed in the
azimuthal direction around the ring. (Such an electric field can be
generated by 
{slowly varying} a magnetic flux through the hole of the annulus.)

The (linear) current response to the applied electric field is defined
to be the conductivity.  A well known result in the literature gives
the so-called intrinsic contribution to the anomalous
Hall effect as the integral of the Berry curvature $\Omega({\bf k})$
over occupied states\cite{AHERMP}.  Here we give the result in two
dimensions (so $\Omega$ is a scalar) for simplicity,
\begin{equation}
\sigma_{xy}^{\rm int}(\mu) = 
e^2 \int \frac{d^2k}{(2 \pi)^2} \,\,  \Omega({\bf k})  \, n_F( \varepsilon({\bf k})-\mu), \label{eq:hall}
\end{equation}
where $n_F$ is the Fermi occupation factor.  If multiple bands are
partially filled, these must be summed over.  

This intrinsic response fully describes the Hall conductivity in the
absence of interactions and absence of disorder (taking a limit of
frequency going to zero, and also taking the limit of very small
applied electric field at the same time, see Appendix
\ref{app:elclean}).  One approach to deriving this result is to use
the Kubo formula.  A more transparent approach is to substitute
Eq.~(\ref{eq:motion2a}) into Eq.~(\ref{eq:motion1a}) and integrate the
anomalous velocity term over all filled states.  In any case, because
the Berry curvature is geometry dependent, so is this resultant Hall
response.

If we would consider the case with disorder, the calculation of the
Hall response would be more complicated but would remain geometry
dependent. In particular, as described for example in
Refs.~\onlinecite{Sinitsyn2007, AHERMP}, there are several
contributions to the Hall conductivity in addition to the intrinsic
part. However, in the limit $\omega \tau \gg 1$ with
  $\omega \rightarrow 0$, the intrinsic part, Eq.~(\ref{eq:hall}),
  correctly gives the Hall conductivity\cite{AHERMP,Nozieres, Dugaev}.
Here $\tau$ is the transport (scattering) lifetime.
Further, when $\omega$ is strictly zero, (at least for {weak and}
smooth disorder) the intrinsic contribution can still be identified as
it is the only piece that reflects the Berry curvature deep within the
Fermi sea\footnote{Note that the Berry flux through the Fermi sea,
  which controls the Hall response via Eq.~(\ref{eq:hall}), can be
  cast via Stokes' theorem as a Fermi surface property (equal to the
  Berry phase acquired by an electron that traverses the Fermi
  surface)\cite{HaldaneAnomalous}. Nonetheless, this quantity is
  intimately tied to the geometry of the Bloch wave functions
  throughout the Fermi sea, whereas the skew-scattering and side-jump
  contributions to the Hall conductivity that result from scattering
  off of impurities are sensitive predominantly to the wave functions
  at the Fermi surface. In particular, for smooth disorder, the side
  jump contribution is proportional to the Berry curvature at the
  Fermi surface\cite{SinitsynSmooth}, which can be tuned (even to
  zero) independently of the Berry flux.}, with all other pieces only
sensitive to the Berry curvature near the Fermi
surface\cite{Sinitsyn2007}.

\subsubsection{Comparison}

It is worth noting that {it is commonly expected that} the response of
a system to a chemical potential difference is the same as its
response to an electric field --- both perturbations applying a bias
that drives current.  However, as we have shown here, in some
circumstances, and depending on the precise questions
  asked, they give different results. One might ask whether the
situation is different in the presence of disorder.  While disorder
makes the calculations of the current response much more complicated,
we can nonetheless be assured that the two situations described above,
with (i) chemical potential difference and (ii) electric field, will
give different Hall current responses, as the former is geometry
independent and the latter is generically geometry dependent.

It is interesting to note that in the geometry independent case (i)
  where we apply a chemical potential difference, the perturbation to
  the system {\it does not change the Hamiltonian of the system at
    all}, it only changes the chemical potential of the particles in
  the leads.  On the contrary the applied electric field in case (ii)
  perturbs the Hamiltonian of the entire system.

\subsection{Thermal Hall Current Response}

Analogous to the two cases in the last subsection, here we consider a
temperature difference or gradient applied in one direction, while the
thermal current orthogonal to the gradient is measured.  The physics
of this heat current response is similar in spirit to that of the
electric current response.

\subsubsection{Geometry Independent Response}

\label{subsub:thermalindep}

Analogous to section \ref{subsub:electricindep} we imagine attaching
our (disorder-free) 
system in a cylindrical geometry to two thermal reservoirs which have a
temperature difference.  We then measure the heat current
perpendicular to the direction between these reservoirs.  We assume
that there is no coupling to phonons or any other degrees of freedom
outside of the system (except coupling to the reservoirs).

Again we use the same tight-binding model, Eq.~(\ref{eq:Hhopping}), which
is completely geometry independent.  Nowhere in the definition of the
model or in the coupling to the reservoirs did we need to specify
the position of the orbitals in the unit cell.  Again we even may add
geometry independent disorder and interactions as in Eqs.~(\ref{eq:Hdis})
and (\ref{eq:Hint}), and the heat current that flows around the circumference of the cylinder must remain independent of the geometric information about the orbital positions within the unit cell.

\subsubsection{Geometry Dependent Response}
\label{subsub:thermaldep}

Analogous to section \ref{subsub:electricdep} here we apply a {\it
  uniform temperature gradient} to the system.  In other words, at
position ${\bf r}_i$ in the sample there will be a weak coupling to a
reservoir with temperature $T({\bf r}_i)$ where $T$ has a uniform
gradient.  Since the reservoir temperature coupled to orbital $i$ is
dependent on the position of orbital $i$, we expect the response to
be explicitly geometry dependent.

This thermal Hall conductivity (also known as the Righi-Leduc
coefficient) can be calculated in several ways. A detailed derivation
is given in Ref.~\onlinecite{ThermalHall} (see also
Ref. \onlinecite{HaldaneAnomalous}) yielding an intrinsic contribution
to the thermal Hall conductivity given by
\begin{eqnarray*}
  \kappa_{xy}^{\rm int}(\mu)  &=&  \frac{-1}{e^2 T} \int d\epsilon \, (\epsilon - \mu)^2 \, \sigma_{xy}^{\rm int}(\epsilon)
   \,                      n_F'(\epsilon - \mu)\\
  & \approx & \frac{\pi^2}{3} \frac{k_B^2 T}{e^2} \sigma_{xy}^{\rm int}(\mu). 
\label{eq:kxyweidemann}
\end{eqnarray*}
In going to the second line, which shows the Wiedemann-Franz relation,
we have assumed the temperature is low and the chemical potential is not at a
singular point (such as at a Dirac node).
The geometry-dependence of the intrinsic Hall conductivity $\sigma_{xy}^{\rm int}(\epsilon)$, inherited from that of the Berry curvature $\Omega({\vec{k}})$, see Eqs.~(\ref{eq:omshift}) and (\ref{eq:hall}), is thus manifested in the thermal Hall conductivity.

To reiterate, as in the case of application of an electric field, the
key distinction between the cases of thermal transport considered here is that the perturbing field %$\phi({\bf r})$ or
$T({\bf r})$ varies smoothly in space and the value of this field felt
by an electron in orbital $i$ depends on that orbital's spatial position.

\subsection{Which responses are measured in experiments?}

In both the electrical and thermal transport cases discussed above, we described two types of
responses that seem as if they should be very similar, but generally
give different results --- one giving a geometry
independent result and the other giving a geometry dependent result.
It is then natural to ask which of these responses is actually
measured in an experiment.  

\begin{itemize}
\item For the electrical current response, in order to measure the
  geometry independent result one would need to work with neutral
  fermions so that they do not feel the electromagnetic field produced
  either by the contacts or by the density of fermions elsewhere in
  the sample.  While this is not the case for electrons, a recent
  experiment with cold fermionic atoms~\cite{Krinner2015} demonstrates
  that it is possible to measure transport of neutral fermions
  subjected to applied chemical potential differences between
  reservoirs. We note that while current densities
  can depend microscopically on details of the geometry, if one measures the total current
  across a defined line cutting across a system, this total current must be
  geometry independent (provided that the pattern of bonds intersected by the cut is held constant, or that the current is evaluated in steady state).

\item For most electronic systems, the natural current response to
  measure will be the usual geometry-dependent
  conductivity\cite{AHERMP,Sinitsyn2007}, since the electric field
  will generally be nonzero.  In particular this implies an intrinsic
  piece of the response given by Eq.~(\ref{eq:hall}).
  This result is believed to hold very generally at least
    in the limit $\omega \tau \gg 1$ and $\omega \rightarrow 0$.

\item For the case of thermal Hall response, to measure the geometry
  independent effect, one must be in a temperature regime where
  phonons are completely decoupled from the device --- as we do not
  want the phonon bath to act as an additional reservoir.  We then can
  attach thermal reservoirs only to the ends of the sample to apply a
  temperature difference.  While such a complete decoupling is only
  obtainable at extremely low temperatures in most electronic systems,
  it is quite easily achieved in cold atom experiments.
  We also note that a system with Coulomb interactions
    between the fermions will not generally be fully geometry
    independent since the Coulomb interaction is sensitive to the physical
    distance between orbitals [i.e., it does not conform to the requirements on interactions discussed around Eq.~(\ref{eq:Hint})].  Nonetheless, it may be the case that
    one can ignore the Coulomb interaction in certain situations so
    long as no density imbalance builds up.
    
\item To measure the usually considered (and geometry dependent)
  thermal Hall response one wants to couple to a thermal reservoir
  throughout the entire sample to impose a uniform gradient.  For a two dimensional system (such as
  graphene) this could be done by sandwiching the sample between two
  reservoirs which themselves have thermal gradients.  It is also
  possible that the system's own phonons can act as this reservoir.

\end{itemize}

\subsection{Thermal transport in spin systems}

The approach of analyzing (thermal) Hall transport is fairly general
and can also be applied to systems with (bosonic) magnon
excitations\cite{ThermalSpin,Lee,Lee2,ThermalSpin0,ThermalSpin2,ThermalSpin3}.
A fairly detailed discussion of this approach is given in
Ref.~\onlinecite{ThermalSpin}.  In such a scheme a band structure is
determined for magnons, and the corresponding Bloch eigenstates can be
used to calculate a Berry curvature.  The thermal Hall conductance is
then given by an integral over the Brillouin
zone\cite{ThermalSpin,ThermalSpin0} (written here in the
two-dimensional case for simplicity)
\begin{equation}
\kappa_{xy} = \frac{-k_B^2 T}{\hbar} \int \frac{d^2k}{(2 \pi)^2} \, c(\omega_{\bf k})  \Omega({\bf k}), \label{eq:thermalspin}
\end{equation}
where $c(\omega_{\bf k})$ is a known simple
function\cite{ThermalSpin,ThermalSpin0} of the frequency
of the magnon,
$\omega_{\bf k}$, and $\Omega(\vec{k})$ is the Berry curvature.  As with the
case of the thermal Hall coefficient obtained from electronic band structure in
section \ref{subsub:thermaldep}, the derivation of this result assumes
coupling to a bath with a uniform thermal gradient.  As with that
result, this formula is manifestly geometry dependent.

As mentioned in the prior section, while coupling to a thermal bath of
uniform thermal gradient may be an accurate representation of certain
experiments, it is also possible that thermal transport experiments
can be constructed where this is not the case.  In particular, if one
has a spin system where only the spins carry heat (meaning no phonons
or other thermal bath), then energy is transported between {\it sites}
according to the geometry-independent graph of exchange couplings that
plays the analogous role to the tight-binding Hamiltonian in
Eq.~(\ref{eq:Hhopping}). In this case, the thermal transport is in
fact geometry-independent, and Eq.~(\ref{eq:thermalspin}) cannot
apply.  While 
  spin systems 
  may host long
  range dipolar interactions in addition to short range exchange
  terms, these dipolar terms are often very weak and may not be
  significant for thermal transport. Thus
  spin systems are
  potentially advantageous for observing geometry independent
  responses.

\section{Further Directions and Conclusions}

In this work, we have investigated the roles of orbital {\it connectivity} and {\it geometry} in electronic lattice systems.
To expose these concepts, we worked within a tight-binding framework where the information about the physical positions $\{\vec{r}_i\}$ of the atomic sites are abstracted away and the electronic dynamics are fully captured by the network of hopping amplitudes and on-site potentials encoded in the tight-binding parameters $\{t_{ij}\}$.
By taking this point of view, new insight can be gained about physical quantities which either do or do not depend explicitly on the {\it geometry} of the system, as encoded in the positions $\{\vec{r}_i\}$.
In particular, we studied how the Bloch band Berry curvature and related observables depend on the atomic positions when the tight-binding parameters are kept fixed. These considerations guided us to a refinement of the ``geometric stability hypothesis'' regarding the stability of fractional Chern insulators, and provided valuable insight into the nature of semiclassical dynamics and various Hall-type transport measurements.

In Sec.~\ref{sec:hall} we examined the Hall response of a
noninteracting electron system with a partially filled band.
Here we discussed the different responses that arise when 
  a chemical potential difference is applied between contacts, or when a uniform electric field is applied to the system. 
We discuss how
the former gives a geometry-independent response, whereas the latter is
geometry-dependent. 
We find a similar situation for
the thermal Hall response where one can either apply a thermal
difference between reservoirs or a coupling to a uniform thermal
gradient over the length of the sample.  We commented that similar
arguments apply to the thermal Hall response due to magnons.  We note
in passing that other responses can also be similarly analyzed, such
as the thermoelectric Nernst or Peltier
responses~\cite{HaldaneAnomalous,Thermoelectric2006}.

While we have worked within the tight-binding formulation, our results are more general.
In principle one could imagine altering a spatial metric while at the same time changing the metric for the continuum Schr\"{o}dinger equation such that the  eigenenergies remain unchanged.   Similarly we will discover a change in Berry curvature and our above arguments will remain unchanged.  (Another way to understand this would be to take a continuum limit of the discrete hopping models that we considered.)

More generally, any calculation can be tested against the backdrop of the type of geometric invariance that we have introduced.  
As with gauge invariance (or covariance), a calculation that fails to transform correctly must be  incorrect at some level.   We propose that this approach can be %of great use
valuable for the field in providing a new type of consistency check for a range of results. 

{\bf Acknowledgements:} The authors acknowledge helpful conversations with
Netanel Lindner, Nick Read, Benoit Doucot, Nigel Cooper, Titus
Neupert, Andrei Bernevig, Gunnar M\"{o}ller, Hannah Price, Justin
Song, and Oded Zilberberg.  SHS has been supported by the Niels Bohr
International Academy, the Simons foundation, and EPSRC Grants
EP/I031014/1, EP/N01930X/1, and EP/S020527/1.  Statement of
compliance with EPSRC policy framework on research data: This
publication is theoretical work that does not require supporting
research data.  MR gratefully acknowledges the support of the Villum
Foundation, and the European Research Council (ERC) under the European
Union Horizon 2020 Research and Innovation Programme (Grant Agreement
No.~678862).

\appendix

\section{Geometric Stability Hypothesis}
\label{app:geometric}

In addition to the Berry curvature, another interesting measure of the ``quantum geometry'' of a band structure is the so-called Fubini-Study metric, defined as\cite{Marzari1997, Roy,Neupert, Dobardzic2013,FubiniStudy1,FubiniStudy2}
\begin{equation} \label{eq:Fubini}
 g_{\mu \nu}({\bf k}) = \frac{1}{2} \left[ \left\langle \frac{\partial  u}{\partial k_\mu} | 
 \frac{\partial u }{\partial k_\nu} \right\rangle - {\cal A}_{\mu} {\cal A}_{\nu}  \right]  + (\mu \rightarrow \nu), 
\end{equation}
where both $u$ and $\cal A$ are functions of $\bf k$, and $\mu$ and $\nu$ run over the Cartesian directions of the system.   
Under a change in geometry as described in Eq.~(\ref{eq:move1}) and the resulting Eq.~(\ref{eq:move2}), this metric transforms nontrivially.   
In the case of two-dimensional bands, two additional interesting quantities derived from this metric are\cite{Roy,Bauer,Jackson2015}:
\begin{eqnarray}
\label{eq:Dk}
 D({\bf k}) &=& \det g(\vec{k}) - \frac{1}{4} |\Omega(\vec{k})|^2 \\
 T({\bf k}) &=& {\rm Tr} \, g(\vec{k}) -  |\Omega(\vec{k})|.
 \label{eq:Tk}
\end{eqnarray}
Note that both $D$ and $T$ are zero for a continuum Landau level, and $D$ is also zero for any two-band model\cite{Milo,Jackson2015}.  Further, it was shown\cite{Roy} that both $T$ and $D$ are nonnegative, and that if $T$ is zero everywhere in the Brillouin zone then $D$ is also. 
 
It was argued that in addition to being favored by flat Berry curvature,   large fractional Chern insulator gaps are favored when the Fubini-Study metric $g_{\mu\nu}(\vec{k})$ is uniform in the Brillouin zone, as well as having $T$ and $D$ close to zero throughout the zone\cite{Dobardzic2013, Roy,Bauer,Jackson2015} [see Eqs.~(\ref{eq:Fubini})-(\ref{eq:Tk})].    These three conditions --- (i) flatness of Berry curvature $\Omega$ (ii) flatness of $g_{\mu\nu}$ and (iii) minimal values of $D$ and $T$ --- are designed to make the band resemble a continuum Landau level as much as possible.   The statement that %obeying these conditions is correlated with
optimizing these conditions maximizes the
many-body gap has been termed the ``geometric stability hypothesis''\cite{Jackson2015} and  has been supported in a number of numerical works\cite{Bernevig,Bauer,Jackson2015}.  

As with the Berry curvature, the Fubini-Study metric, as well as $D$ and $T$, generically change under geometric transformations that leave the hoppings and interaction matrix elements (in the site basis) completely unchanged. Crucially, under such transformations, the many-body gap is unchanged.   Thus, the hypothesis cannot be correct as  stated.   Nonetheless, one cannot discard the observed correlations between gaps and various band parameters [hypotheses (i)-(iii) above] which have been measured in prior numerical works\cite{Bernevig,Bauer,Jackson2015}.

We note that in several prior works, the geometry (i.e., the positions of sites in the unit cell)  used in numerical simulations happens to correspond to the geometry that minimizes the variance of Berry curvature over the Brillouin zone (i.e., it maximizes the flatness), for a given set of tight-binding parameters.  For example, in the fractionally filled Haldane model studied numerically in Refs.~\onlinecite{Bernevig,Jackson2015} the minimal variance of the Berry curvature  (i.e., the highest ``flatness'') happens to occur when the orbitals lie on a simple honeycomb, which is the geometry that was considered in Ref.~\onlinecite{Jackson2015}.    Similarly, for the Hofstadter model with small flux per plaquette, the Berry curvature is most flat when the orbitals in the magnetic unit cell lie equally spaced, which is exactly the case studied in Ref.~\onlinecite{Bauer}.

Inspired by this coincidence, we now suggests a possible modification of the geometric stability hypothesis which would 
be consistent with the geometric invariance required by the arguments in the main text. We propose to measure the variation (the flatness) of the Berry curvature (and Fubini-Study metric) only after optimizing over all possible geometries of the orbitals.  In other words, given a particular microscopic model, we accept that the Berry curvature is a function of the geometry of orbitals in the model, but the many-body gap [for Hubbard-like interactions as in Eq.~(\ref{eq:Hint})] is not.   For a fixed geometry, we measure the variance of the Berry curvature; then we vary over {\it all possible} geometries of the orbitals (all positions of the orbitals in the unit cell), while keeping the tight-binding parameters in the Hamiltonian fixed.  We then focus on the particular geometry for which the curvature is most flat.  We suggest that the many body gaps will correlate with this optimized quantity.  Indeed, this modified conjecture is supported by previous numerical data from Refs.~\onlinecite{Bauer,Jackson2015} which did focus on this optimized orbital geometry.    One can apply similar reasoning to the Fubini-Study metric and the resulting quantities $D$ and $T$ as well [Eqs.~(\ref{eq:Fubini})-(\ref{eq:Tk})].  This prescription is well-defined, and consistent with the geometric invariance principle.
It is a matter for future research to decide if this prescription does indeed predict many body gaps accurately  (and if so, why). 

\section{Explicit Calculations of the Hall Current Response}
\label{app:current}

In this appendix, we provide two proof-of-principle, explicit Hall
transport calculations to support the discussion of geometry-dependent
and independent responses in the main text.  For simplicity we will
assume a uniform two dimensional system with
noninteracting fermions, in a system free of any disorder.  (We take
this special case just for demonstration -- our qualitative
conclusions about geometry-dependence and independence do not rely on
the absence of disorder.)  {We also note that here we have not
  provided any mechanism for electrons to scatter and re-equilibrate.}
Although there are several methods for calculating the Hall response,
our approach, in the spirit of Boltzmann theory, will be to consider
the dynamics of wave packets in phase-space.  The particle current
density is given by
\begin{equation}
{\bf j}= \int \frac {d{\bf k}}{(2 \pi)^2}  \, n({\bf k})\, \dot {\bf r}({\bf k}),
\label{eq:jex0}
\end{equation}
where $n({\bf k})$ is the occupancy of the state ${\bf k}$.  From
Eq.~(\ref{eq:motion1a}), there are two contributions to
$\dot {\bf r}(\vec{k})$: the regular (i.e., group) velocity
${\bf v}_g({\bf k}) = \nabla_{\bf k} \varepsilon({\bf k})$, and the
anomalous velocity
${\bf v}_a(\vec{k}) = \dot {\bf k} \times \Omega(\vec{k})$.  In
principle $n({\bf k})$ and $\dot {\bf r}({\bf k})$ may be functions of
position ${\bf r}$, %as well giving us
giving a current density ${\bf j}(\vec{r})$ that is a function of
position as well.  In the presence of inhomogeneities one may also
obtain magnetization currents\cite{NiuReview}.  However, 
the simple demonstrations here avoid this added complexity
since we have chosen to consider cases which are spatially uniform.

\subsection{Applied Electric Field, Clean Limit}
\label{app:elclean}

{We consider an infinitely large (or periodic),
  disorder-free system.  We will apply a weak time-dependent uniform electric
  field ${\bf E}(t)$ at low frequency.  
The system starts in a thermal occupation
$n^{(0)}({\bf k}) = n_F(\varepsilon({\bf k}) - \mu)$, and we adiabatically
turn on the electric field~\footnote{Although the system may be
  gapless there is still a notion of adiabaticity since, in the absence
  of disorder and interactions, momentum conservation forbids any
  transitions to other low energy states in linear response. For small $\vec{E}$ and $\omega$, interband transitions are also suppressed.}.
In this case the ${\bf k}$ states simply
accelerate freely via $\dot {\bf k} = -e {\bf E}(t)$, as in
Eq.~(\ref{eq:motion2a}), without scattering or relaxing.
The occupation distribution therefore adiabatically evolves as $n(\vec{k}, t) = n_F(\varepsilon[\vec{k}(t)])$, with the time-dependent wave vector 
\begin{equation}
  \label{eq:koft}
{\bf k}(t) = {\bf k} - e \int^t_{-\infty} dt' \,\, {\bf E}(t').
\end{equation}
Using Eq.~(\ref{eq:jex0}), the resulting current {density} is given by 
\begin{equation}
{\bf j}(t) = \int \frac{d{\bf k}}{(2 \pi)^2} \, n_F(\varepsilon[{\bf k}(t)] - \mu)\, \dot {\bf r}({\bf k}). 
\label{eq:jex}
\end{equation}

In the absence of ${\bf E}$, the current density in the uniform system must
vanish everywhere. %be zero.
We are interested in the effects of ${\bf E}$ at linear order.  There
are two places in Eq.~(\ref{eq:jex}) where $\vec{E}$ enters: (i) in
the anomalous velocity part of $\dot {\bf r}$, $\vec{v}_a(\vec{k})$,
and (ii) in ${\bf k}(t)$.  We will consider these two contributions
separately, to linear order in ${\bf E}$.  To stay in the linear
response regime, we therefore assume that the frequency is low enough
to be in the adiabatic limit with respect to interband transitions,
while the electric field is also small enough that
$e |{\bf E}|/\omega$ is small enough that nonlinear contributions in
$\vec{E}$ can be neglected.
Such nonlinear corrections (of order ${\bf E}^2/\omega$)  are discussed for example in Ref.~\onlinecite{SodemannFu}.
{Note that in cases where there is an equilibration mechanism, such as phonons or electron-electron interaction,  the current calculation only remains correct so long as the time scale for this equilibration is {much} longer than $1/\omega$. }

Considering contribution (i) defined above, in linear response we obtain a contribution to the Hall current {density} from the anomalous velocity term given by
\begin{equation}
 {\bf j}_H^0(t) =  e {\bf E}(t) \,  \times \hat {\vec{z}} \int \frac{d{\bf k}}{(2 \pi)^2}\, \Omega({\bf k})\, n^{(0)}({\bf k}),  \label{eq:jH0}
\end{equation}
where $\hat{\vec{z}}$ is the normal to the plane.   This thus gives a low-frequency Hall conductivity of the form in Eq.~(\ref{eq:hall}).    

We now consider contribution (ii).  Here Eq.~(\ref{eq:koft}) gives the dependence of $\vec{k}(t)$ on $\vec{E}$.
Expanding for small ${\bf E}$ and using $\vec{v}_g(\vec{k}) = \nabla_{\vec{k}}\varepsilon(\vec{k})$, for Eq.~(\ref{eq:jex}) we obtain
\begin{equation}
  %\varepsilon\Big[{\bf k} - e \int^t_{-\infty}\!\!\!\!\!\! dt'\, {\bf E}(t')\Big]
 \varepsilon\big[{\bf k}(t)\big] = \varepsilon({\bf k}) -e {\bf v}_g({\bf k}) \cdot \int^t_{-\infty}\!\!\!\! dt'\,{\bf E}(t') + \cdots. \label{eq:epsexp} 
\end{equation}
Substituting this expression into Eq.~(\ref{eq:jex}) and expanding to linear order in $\vec{E}$ we obtain a contribution to the current {density}
\begin{equation}
  \nonumber
  \label{eq:delta_j}\delta  j_\mu(t) = -e \!\!\int_{-\infty}^t \!\!\!\!\!\!\! dt' \, E_\nu(t') 
  \!\! \int \!\!\frac{d{\bf k}}{(2 \pi)^2} \, v_{g,\mu}({\bf k})  v_{g,\nu}({\bf k})\,\,  n_F'(\varepsilon({\bf k}) - \mu),
\end{equation}
where $\mu,\nu$ label the Cartesian directions.
Crucially, $\delta  j_\mu(t)$
% in Eq.~(\ref{eq:delta_j})
gives no Hall component of the response, as the $\vec{k}$-integral on the right hand side is symmetric between $\mu$ and $\nu$.
(Note that one must take only the antisymmetric component $\sigma_{xy} - \sigma_{yx}$ to isolate the Hall component.)
Thus the Hall current {density} remains that given by Eq.~(\ref{eq:jH0}).  This response is geometry dependent due to the geometry dependence of $\Omega(\vec{k})$.

\subsection{Chemical Potential Difference, Clean Limit}
\label{app:chemclean}

Here we consider the case of a system of neutral fermions being
coupled to two reservoirs at different chemical potentials $\mu_1$ and
$\mu_2$.  In this case, the dynamics are fully determined by the
hopping of particles between sites of the unperturbed tight-binding
network, and the response of the system must be geometry independent.

The situation we envisage is inspired by the
  recent experiments in Ref.~\onlinecite{Krinner2015}.  We assume that
  the reservoirs are separated in the $\hat{\vec{x}}$ direction; it is
  most convenient to think about the sample as being periodic in the
  $\hat{\vec y}$ direction and infinitely long in the $\hat {\vec x}$
  direction so that we are considering an infinitely long cylinder all constructed of the same material (the
  same hopping model).  Initially, we split the cylinder into three
  disconnected pieces, $x < 0$ the left reservoir, $0 < x < L$ the
  system, and $x > L$ the right reservoir.  We fill the left and right
  reservoirs with fermions such that $\mu_1 \neq \mu_2$.  Then at some
  time we connect the three systems together, and we let current flow
  between the reservoirs. In the long time limit there will be a
  steady state current flow in the system, which we measure.  The
  steady state current flow will be independent of details such as the
  initial state in the system section $0 < x < L$, and the precise procedure
  for connecting up the three pieces. }

In the clean limit, where there is no scattering, transport
between the two reservoirs is essentially ballistic.  Since there is
no electric field, there is no anomalous velocity. The total current
density within the bulk is then 
\begin{equation}
 {\bf j} = \int \frac{d{\bf k}}{(2 \pi)^2}  \nabla_{\bf k}  \varepsilon({\bf k})  \,n({\bf k}),
\label{eq:chempotentialcurrent}
\end{equation}
where the distribution is given by
$$
 n(\vec{k}) = \left\{ \begin{array}{lll} n_F(\varepsilon({\bf k}) - \mu_1) & &   v_x({\bf k}) > 0 \\ 
 n_F(\varepsilon({\bf k}) - \mu_2) & &   v_x({\bf k}) < 0.  \end{array}\right. 
$$
If one considers a dispersion $\varepsilon({\bf k})$ which is
symmetric under $k_y \leftrightarrow -k_y$, the Hall current density
will be zero, independent of the Berry curvature of the system's bands
and of the orbital geometry.  For a less symmetric
dispersion 
the result may be nonzero (but still independent of the Berry
curvature).

\bibliography{refs.bib}

\begin{thebibliography}{63}
\bibitem [{not()}]{note:Milo}%
  \BibitemOpen
  \href@noop {} {}\bibinfo {note} {An interesting possibility, suggested by
  Ref.~\onlinecite{Milo}, is that one should focus on special points in the
  Brillouin zone where the Berry curvature is independent of geometric
  transformation. This appears possible for two-band models only.}\BibitemShut
  {Stop}%
\end{thebibliography}

\end{document}